# Mobility and Threshold Voltage Extraction in Transistors with Gate-Voltage-Dependent Contact Resistance

Robert K. A. Bennett,[1] Lauren Hoang,[1] Connor Cremers,[1] Andrew J. Mannix,[2] and Eric Pop[1,2,3*]

[1]*Department of Electrical Engineering, Stanford University, Stanford, CA 94305, U.S.A.*

[2]*Department of Materials Science and Engineering, Stanford University, Stanford, CA 94305, U.S.A.*

[3]*Department of Applied Physics, Stanford University, Stanford, CA 94305, U.S.A.*

[*]Contact: epop@stanford.edu

This document is the authors' version of a manuscript published in *npj 2D Materials and Applications*. The published article can be accessed at https://doi.org/10.1038/s41699-024-00506-4 or https://www.nature.com/articles/s41699-024-00506-4.

The full citation for the published article is:

> Bennett, R.K.A., Hoang, L., Cremers, C., Mannix, A.J., Pop, E. "Mobility and threshold voltage extraction in transistors with gate-voltage-dependent contact resistance," *npj 2D Materials and Applications* **9**, 13 (2025). DOI: 10.1038/s41699-024-00506-4

**Abstract:** The mobility of emerging (e.g., two-dimensional, oxide, organic) semiconductors is commonly estimated from transistor current-voltage measurements. However, such devices often experience contact gating, i.e., electric fields from the gate modulate the contact resistance during measurements, which can lead conventional extraction techniques to estimate mobility incorrectly even by a factor >2. Although this error can be minimized by measuring transistors at high gate-source bias $|V_{gs}|$, this regime is often inaccessible in emerging devices that suffer from high contact resistance or early gate dielectric breakdown. Here, we propose a method of extracting mobility in transistors with gate-dependent contact resistance that does not require operation at high $|V_{gs}|$, enabling accurate mobility extraction even in emerging transistors with strong contact gating. Our approach relies on updating the transfer length method (TLM) and can achieve <10% error even in regimes where conventional techniques overestimate mobility by >2×.

**Introduction:** The electron and hole mobilities of emerging semiconductors are frequently estimated from measured current vs. voltage characteristics (e.g., from drain current $I_d$ vs. gate-source voltage $V_{gs}$) of field-effect transistors (FETs).[1] Many such transistors have contact resistance that is a function of gate voltage due to electrostatic gate fields that affect the energy barrier at the contact-channel interface.[2,3] This *contact gating* effect is often associated with back-gated FETs (**Figure 1a**), where the back gate can directly modulate the carrier density at the contact (**Figure 1b**). However, recent work has shown that top-gated FETs (**Figure 1c**) can also electrostatically control the contacts at their edges **(Figure 1d**).[4,5]

Further complicating matters, the channel resistance $R_{ch}$ and contact resistance $R_C$ of contact-gated FETs often change at different rates (**Figure 1e**), causing these devices to exhibit two apparent threshold voltages: one associated with channel turn-on, and another dictated by contact turn-on (**Figure 1f**).[6] When the channel turns on before the contacts (i.e., at lower $V_{gs}$ in *n*-channel FETs, as in Figures 1e,f), $I_d$ is limited by $R_C$ and can remain low even when the channel is fully turned on (Figures 1e,f, **Region I**). As $V_{gs}$ increases, the contacts begin to turn on (Figures 1e,f, **Region II**) before the device eventually reaches a channel-dominated regime (Figures 1e,f, **Region III**), leading to a distinct kink[7] in the $I_d$ vs. $V_{gs}$ characteristics associated with the transition between the contact- and channel-limited regimes.

Because $I_d$ is contact-limited or contact-influenced in Region II of Figure 1f, both $I_d$ and the transconductance ($g_m = \partial I_d/\partial V_{gs}$) here are dominated by the contacts rather than by the channel. Thus, attempting to estimate the channel mobility ($\mu$) using the conventional linear extrapolation method (i.e., asserting $\mu \propto g_m$) in this region can result in severe $\mu$ overestimation[3,6-9] when $R_C$ dominates and decreases as $|V_{gs}|$ increases. (In the special case where $R_C$ remains constant as $V_{gs}$ increases, the mobility can be underestimated instead.) Therefore, $\mu$ should instead be extracted from the slope of Region III in Figure 1f (where devices are channel-limited).[7,8] However, this approach is often infeasible for emerging semiconductor devices whose large $R_C$ and/or early gate dielectric breakdown can make this high-$V_{gs}$ region hard to reach experimentally. Furthermore, when Region III of the $I_d$ vs. $V_{gs}$ curve is inaccessible before dielectric breakdown (e.g., due to large $R_C$ and/or high threshold voltage $|V_T|$), the $I_d$ vs. $V_{gs}$ curve may show only a single linear region simply because the $V_{gs}$ sweep ends early. For this reason, it can even be challenging to establish if a device is channel- or contact-limited based on its transfer characteristics.[10]

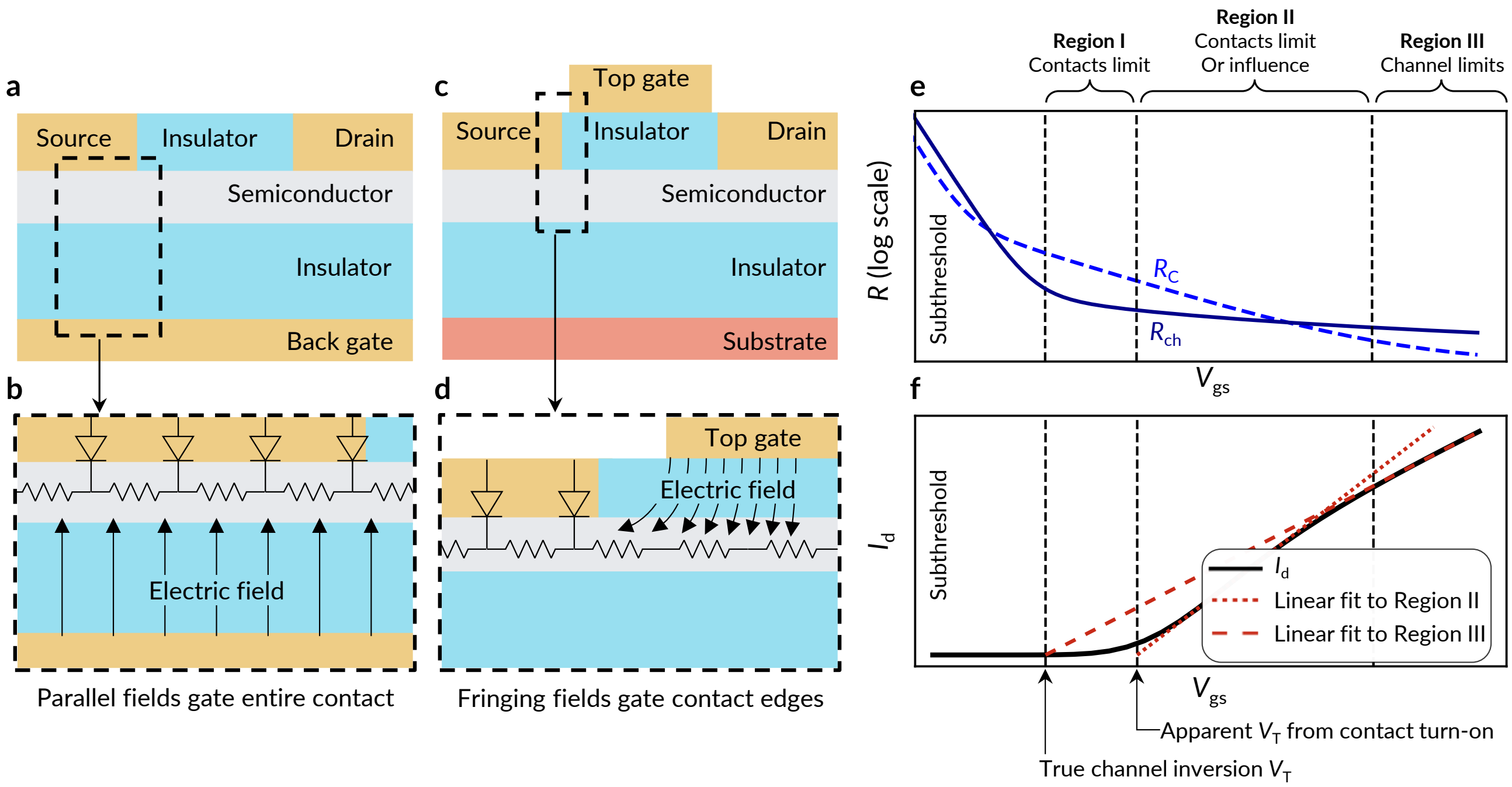

**Figure 1**: **Contact gating in field-effect transistors (FETs). (a)** Schematic of a back-gated FET, where inset **(b)** shows the parallel field from the back gate directly gating the entire contact. **(c)** schematic of a top-gated FET, where inset **(d)** shows fringing fields gating the contact edge. (Top-gated devices without underlaps would have their contact edges gated directly by the parallel field instead.) As a result of contact gating, the channel resistance $R_{ch}$ and contact resistance $R_C$ can decrease at different rates as $V_{gs}$ increases (**e**), leading to a distinct kink in the resultant $I_d$ vs. $V_{gs}$ characteristics **(f)**. After exiting the subthreshold region, the channel may turn on even as the contacts remain off, suppressing $I_d$ (e,f, **Region I**). The contacts then turn on as $V_{gs}$ further increases, leading to a sharp increase of $I_d$ (e,f, **Region II**). The magnitude and slope of $I_d$ here are dictated by the contacts rather than the channel; attempting to extract the channel mobility from this region with conventional techniques can result in severe overestimation. Finally, the FET returns to a channel-limited regime when $R_C < R_{ch}$ (e,f, **Region III**); $\mu$ can usually be safely extracted from this region.

To avoid $\mu$ overestimation due to contact gating, researchers can use four-terminal geometries to directly probe and subtract the voltage drop across the contacts.[9,11] However, care must be taken to ensure that the voltage probes are entirely non-invasive, which can be difficult in practice.[12-15] The Y-function method[16] can also correct for mild contact gating[17] but relies upon accurate $V_T$ extraction, making this method unsuitable for devices that cannot access Region III in Figure 1f.[6] We demonstrate later in this work that the transfer length method (TLM) approach[1] is similarly unreliable for extracting $\mu$ of contact-gated FETs.

In this work, we propose a method of extracting the channel $\mu$ and $V_T$ of transistors that remains valid even for strongly contact-gated devices. This approach takes inspiration from the conventional TLM method[1] and can be used to analyze families of two-terminal devices that cannot access the channel-limited regime (Figure 1e,f, Region III) in their $I_d$ vs. $V_{gs}$ measurements. We validate our proposed method using synthetic data generated by a technology computer-aided design (TCAD) simulator[18] and find that it accurately extracts $\mu$ even for devices where conventional methods overestimate $\mu$ by 2-3×, enabling accurate $\mu$ and $V_T$ extraction in devices with strong contact gating.

**Results:** Our proposed extraction is summarized in **Figure 2** and explained in detail below. We provide Python code to automate this extraction in a GitHub Repository[19] and in **Supplementary Section 4**. Additionally, we provide an accompanying tutorial for this code in **Supplementary Section 1.**

**Model derivation:** Here, we treat a contact-gated FET as a channel between gate-voltage-dependent source and drain resistors ($R_s$ and $R_d$ respectively), as shown in **Figure 2a**.[20] The intrinsic gate-to-source and drain-to-source biases (after considering the voltage drops across $R_s$ and $R_d$) are $V_{gs}' = V_g - V_s'$ and $V_{ds}' = V_d' - V_s'$, where $V_s'$ and $V_d'$ are defined in Figure 2a. In the linear region ($V_{gs}' > V_T$ and $V_{ds}' < V_{gs}' - V_T$), $I_d$ is:

$$\frac{I_d}{W} = \frac{\mu C_{ox}}{2L_{ch}}\left[2\left(V_{gs}' - V_T\right)V_{ds}' - {V_{ds}'}^2\right] \qquad (1)$$

where $C_{ox}$ is the oxide capacitance and $W$ and $L_{ch}$ are the width and length of the channel. In this extraction, we build a system of equations based on eq. (1) that we use to simultaneously solve for $\mu$ and $V_T$. In eq. (1), $V_T$ refers explicitly to the true channel $V_T$ associated with channel inversion (as defined in Figure 1f); thus, this channel $V_T$ often cannot be extracted directly in contact-gated devices. [For emerging FETs with intrinsic channels (no counter-doping), such as two-dimensional (2D) FETs, the channel is considered inverted when the carrier concentration is approximately equal to the density of states at the relevant band edge.[21]]

To build our system of equations, we use a TLM-like approach where we consider a family of devices with varied $L_{ch}$ (using multiple two-terminal devices or a larger TLM-like test structure). As $R_s$ and $R_d$ are $V_{gs}$-dependent, we use $I_d$ vs. $V_{ds}$ sweeps at fixed values of $V_{gs}$ to ensure they remain constant. Further, as $R_s$ and $R_d$ contain Schottky diodes, they are nonlinear circuit elements, i.e., their resistances are functions of $I_d$. To ensure constant $R_C$, we therefore perform the extraction at a constant current.

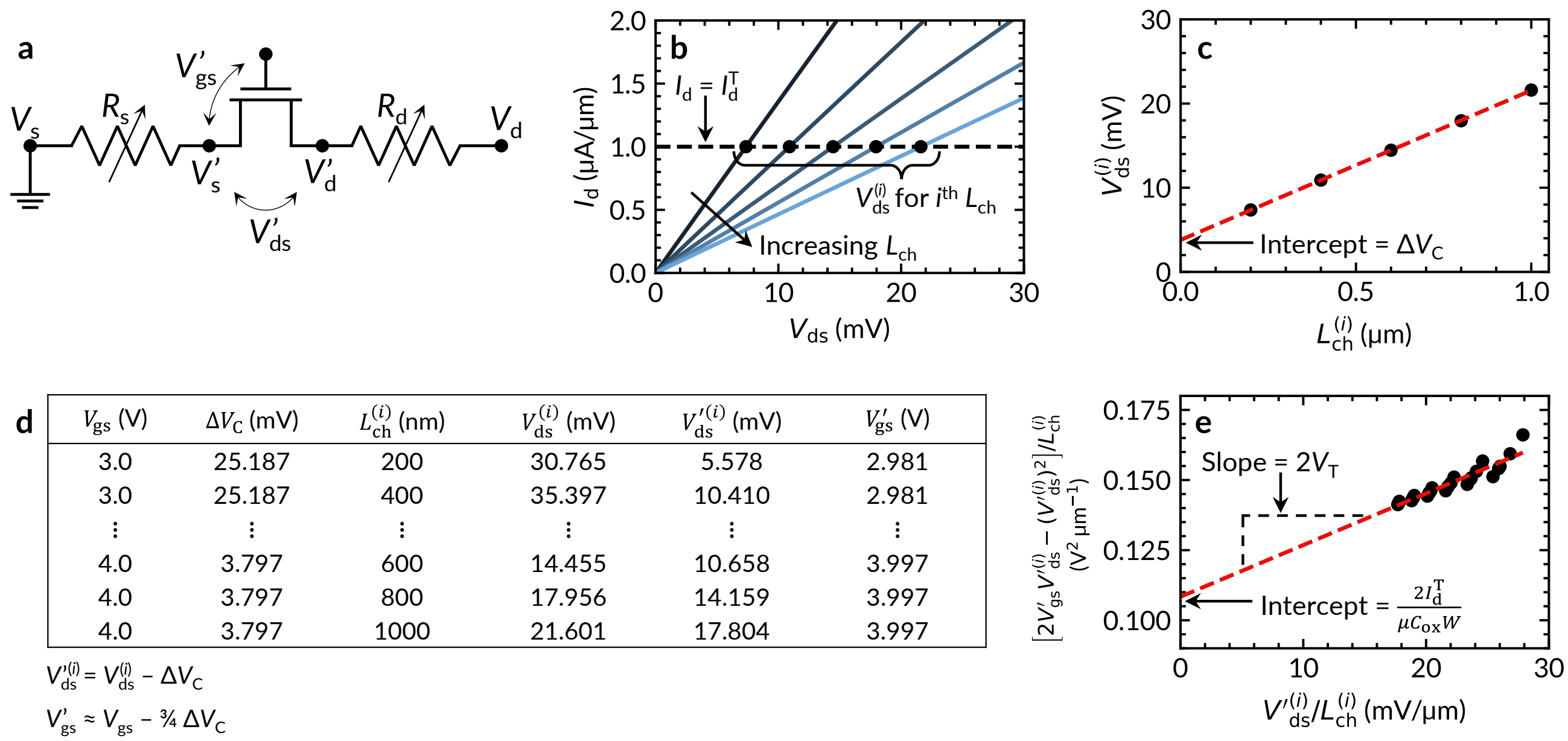

d

| $V_{gs}$ (V) | $\Delta V_C$ (mV) | $L_{ch}^{(i)}$ (nm) | $V_{ds}^{(i)}$ (mV) | $V_{ds}'^{(i)}$ (mV) | $V'_{gs}$ (V) |
|---|---|---|---|---|---|
| 3.0 | 25.187 | 200 | 30.765 | 5.578 | 2.981 |
| 3.0 | 25.187 | 400 | 35.397 | 10.410 | 2.981 |
| ⋮ | ⋮ | ⋮ | ⋮ | ⋮ | ⋮ |
| 4.0 | 3.797 | 600 | 14.455 | 10.658 | 3.997 |
| 4.0 | 3.797 | 800 | 17.956 | 14.159 | 3.997 |
| 4.0 | 3.797 | 1000 | 21.601 | 17.804 | 3.997 |

$V_{ds}'^{(i)} = V_{ds}^{(i)} - \Delta V_C$

$V'_{gs} \approx V_{gs} - ¾\, \Delta V_C$

**Figure 2: Summary of our proposed extraction.** We treat a contact-gated FET as a channel between contact resistors **(a)**. We begin the extraction in **(b)** by performing $I_d$ vs. $V_{ds}$ sweeps at a fixed $V_{gs}$ for a family of devices with varying channel lengths $L_{ch}$. For the $i^{th}$ $L_{ch}$, we record the $V_{ds}$ at which the drain current reaches a target drain current $I_d = I_d^T$ as $V_{ds} = V_{ds}^{(i)}$. We then plot $V_{ds}^{(i)}$ vs. $L_{ch}$ in **(c)** and extrapolate to find the voltage drop across the contacts $\Delta V_C$. We repeat this procedure at multiple $V_{gs}$ to compile the table in **(d)** (using the equations below the table to calculate the intrinsic voltages). Finally, we use data from **(d)** to prepare the plot shown in **(e)**, perform linear regression to find the slope $m$ and $y$-intercept $b$, and extract $V_T = m/2$ and $\mu = 2I_d^T/(bC_{ox}W)$.

With these considerations in mind, we begin by choosing a target drain current $I_d^T$. We then perform $I_d$ vs. $V_{ds}$ sweeps for each $L_{ch}$ at a common $V_{gs}$, recording the $V_{ds}$ at which the device with the $i^{th}$ channel length reaches $I_d = I_d^T$ as $V_{ds} = V_{ds}^{(i)}$ (**Figure 2b**). Then, we plot $V_{ds}^{(i)}$ vs. $L_{ch}$ and perform linear regression (**Figure 2c**); the $y$-intercept of the line of best fit yields the voltage drop across the contacts $\Delta V_C$ at $I_d = I_d^T$. Here, $I_d^T$ must be chosen such that the extracted $V_{ds}^{(i)}$ values are small (all $V_{ds}^{(i)} \ll V_{gs} - V_T$) to minimize the quadratic term in eq. (1); otherwise, the linear fit in Figure 2c becomes invalid. Additionally, a small $V_{ds}^{(i)}$ ensures that the vertical electric field near the drain is similar for all channel lengths, helping to ensure that $R_d$ remains constant across all devices.

Next, we partition $\Delta V_C$ into the voltage drops across the source and drain, $\Delta V_s$ and $\Delta V_d$. As $R_s$ and $R_d$ contain reverse and forward biased Schottky diodes, respectively, we have $R_s > R_d$.[20] Further, as $R_d$ approaches 0 for high drain bias,[22] we have:

$$0 \leq \Delta V_{\mathrm{d}} \leq \Delta V_{\mathrm{C}}/2 \tag{2}$$

$$\Delta V_{\mathrm{C}}/2 \leq \Delta V_{\mathrm{s}} \leq \Delta V_{\mathrm{C}} \tag{3}$$

For simplicity, we take the centers of these ranges, i.e., $\Delta V_{\mathrm{d}} \approx \frac{1}{4}\Delta V_{\mathrm{C}}$ and $\Delta V_{\mathrm{s}} \approx \frac{3}{4}\Delta V_{\mathrm{C}}$, and estimate the true intrinsic voltages as $V_{\mathrm{ds}}'^{(i)} = V_{\mathrm{ds}}^{(i)} - \Delta V_{\mathrm{C}}$ and $V_{\mathrm{gs}}' = V_{\mathrm{gs}} - \Delta V_{\mathrm{s}} \approx V_{\mathrm{gs}} - \frac{3}{4}\Delta V_{\mathrm{C}}$.

We use the above approach to extract $V_{\mathrm{ds}}'^{(i)}$ and $V_{\mathrm{gs}}'$ at multiple fixed $V_{\mathrm{gs}}$ values to compile the table in **Figure 2d**. Next, we use these tabulated values to build a system of equations from which we extract $\mu$ and $V_{\mathrm{T}}$. To do so, we rearrange eq. (1) into:

$$\frac{2V_{\mathrm{gs}}'V_{\mathrm{ds}}' - V_{\mathrm{ds}}'^{2}}{L_{\mathrm{ch}}} = 2V_{\mathrm{T}}\frac{V_{\mathrm{ds}}'}{L_{\mathrm{ch}}} + \frac{2I_{\mathrm{d}}}{\mu C_{\mathrm{ox}} W} \tag{4}$$

As eq. (4) is in the form $y = mx + b$ [with $m = 2V_{\mathrm{T}}$, $x = V_{\mathrm{ds}}'/L_{\mathrm{ch}}$, and $b = 2I_{\mathrm{d}}/(\mu C_{\mathrm{ox}} W)$], we use rows from Figure 2d to plot $(2V_{\mathrm{gs}}'V_{\mathrm{ds}}'^{(i)} - V_{\mathrm{ds}}'^{(i)2})/L_{\mathrm{ch}}^{(i)}$ as a function of $V_{\mathrm{ds}}'^{(i)}/L_{\mathrm{ch}}^{(i)}$ and perform linear regression (**Figure 2e**). We then use the extracted slope and intercept to calculate $V_{\mathrm{T}}$ and $\mu$ from known quantities.

Although we present a derivation for *n*-type devices, this procedure can easily be adapted to *p*-type devices by repeating the derivation starting from the *p*-type analogue of eq. (1). Alternatively, one could apply the above procedure to *p*-type devices by negating the input $V_{\mathrm{gs}}$, taking the absolute value of $V_{\mathrm{ds}}$, and then negating the extracted $V_{\mathrm{T}}$.

We note that $\Delta V_{\mathrm{C}}$ extracted in Figure 2c is accompanied by an associated error that leads to uncertainty in the extracted $\mu$ and $V_{\mathrm{T}}$. Simple analytic techniques cannot propagate this error to the final $\mu$ and $V_{\mathrm{T}}$ because the quantities along both the *x*- and *y*-axes in Figure 2e are error-prone, where the error in *x* and *y* values are not mutually independent. Hence, we instead use the Monte Carlo approach described in **Supplementary Section 2** to propagate this error, allowing us to improve the estimates for the nominal $\mu$ and $V_{\mathrm{T}}$ and their standard errors. This Monte Carlo approach is implemented in Python code that we have uploaded online to a GitHub Repository.[19]

**Model validation:** We validate our proposed extraction by using it to estimate $\mu$ and $V_{\mathrm{T}}$ from current-voltage characteristics generated by Sentaurus Device TCAD.[18] This approach allows us to assess the accuracy of the extraction because $\mu$ and $V_{\mathrm{T}}$ are known *a priori*: $\mu$ is a simulation input parameter, and the channel $V_{\mathrm{T}}$ can be extracted from equivalent devices (simulated in Sentaurus) without contact resistance.

The contact-gated devices in our TCAD simulations have nominal Schottky barrier heights between $\phi_B$ = 0.15 to 0.6 eV. These are "nominal" values because the TCAD simulations include image force lowering (IFL)[23] and tunneling at the contacts; these listed $\phi_B$ are barrier heights before IFL (i.e., the $\phi_B$ we list are the difference between the semiconductor's electron affinity and the metal's work function). All devices are back-gated transistors (Figure 1a) with $HfO_2$ gate insulators (relative dielectric constant $\kappa = 20$ and equivalent oxide thickness EOT = 10 nm). The simulated channel thickness is 0.615 nm, corresponding to monolayer $MoS_2$,[24,25] and the mobility is set to $\mu = 50$ cm$^2$V$^{-1}$s$^{-1}$.

In **Figures 3a-d**, we plot TCAD-generated $I_d$ vs. $V_{ov}$ sweeps (where the overdrive voltage $V_{ov} = V_{gs} - V_T$) at $V_{ds} = 0.1$ V for each $\phi_B$ at $L_{ch}$ = 200, 400, ..., 1000 nm. Devices with $\phi_B \geq 0.3$ eV clearly display the signature kink of contact gating,[7] especially at small $L_{ch}$ (where $R_C$ exceeds the channel resistance at low $V_{ov}$). Next, we plot the extracted $\mu$ and $V_T$ for each $\phi_B$ using our proposed extraction method and three conventional techniques: the linear extrapolation, Y-function, and TLM approaches.[1,16,26] These conventional techniques are applied directly on the synthetic $I_d$ vs. $V_{gs}$ data shown in Figures 3a-d, whereas our extraction uses separate synthetic $I_d$ vs. $V_{ds}$ data. The linear extrapolation and Y-function techniques both use the longest channel device ($L_{ch}$ = 1000 nm), whereas our proposed method and the TLM approach use the full range of $L_{ch}$ = 200 to 1000 nm in Figures 3a-d. We calculate standard error using the method described in Supplementary Section 2 for our method, and we use the standard error from linear regression (equivalent to the 68% confidence interval) for the TLM approach. (We do not include standard error for the linear extrapolation or Y-function approaches because the synthetic data is noiseless.)

In **Figures 3e-h**, we plot the $\mu$ extracted from each method vs. $V_{ov}$/EOT. The horizontal axes in these plots are shown only up to the point where the $\mu$ obtained by the four extraction methods have converged. At $\phi_B$ = 0.15 eV (Figure 3e), we find that all extractions yield reasonable estimates for $\mu$, with a worst-case underestimation of ~15% for the Y-function method. As $\phi_B$ increases, however, we find that conventional methods begin to severely overestimate $\mu$ due to contact gating. We also note the TLM approach predicts a small standard error (<10%) in Figures 3g,h ($\phi_B$ = 0.45 and 0.6 eV) despite overestimating $\mu$ by over 2×. In other words, the standard error estimated from the TLM approach does not accurately reflect the true uncertainty in the extracted $\mu$ when $\phi_B$ is large and $V_{ov}$ is limited (e.g., by early dielectric breakdown). In contrast, our method estimates $\mu$ more accurately than conventional methods, with a worst-case overestimation of ~20% at $\phi_B$ = 0.6 eV and low $V_{ov}$ (Figure 3h), and with the true $\mu$ being captured within our error bars (unlike the TLM method). We note that the TLM approach requires that devices with different $L_{ch}$ be measured at a common carrier density, i.e., at a common $V_{ov}$.[1] In the present work, $V_{ov}$ is referenced with respect to $V_T$ estimated by linear extrapolation; in **Supplementary Section 3**, we study the accuracy of the TLM approach when instead using $V_T$ defined at a constant-current (e.g., 100 nA/μm).

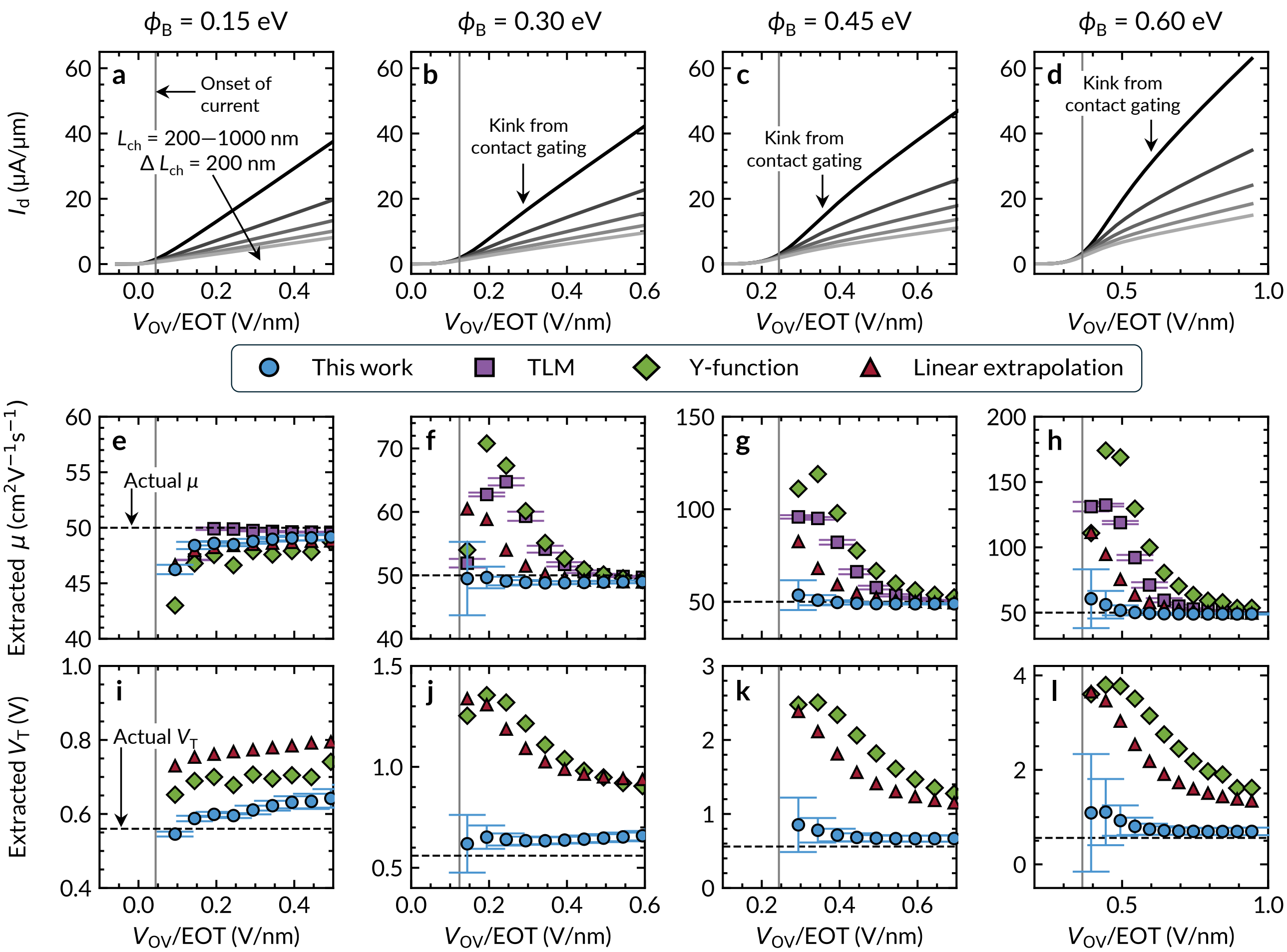

**Figure 3: Validation of our proposed extraction.** $I_d$ vs $V_{ov}$ for families of devices with $L_{ch}$ = 200, 400, ..., 1000 nm and Schottky barrier height $\phi_B$ = **(a)** 0.15 eV, **(b)** 0.3 eV**, (c)** 0.45 eV, and **(d)** 0.6 eV. $x$-axes show different ranges of $V_{ov}$/EOT to highlight device characteristics and extractions near $I_d$ turn-on for different $\phi_B$. Solid gray lines mark the approximate $I_d$ turn-on for each device family. **(e)** The mobility $\mu$ extracted with our method and the linear extrapolation, Y-function, and TLM approaches for device families with $\phi_B$ = 0.15 eV, **(f)** 0.3 eV, **(g)** 0.45 eV, and **(h)** 0.6 eV plotted vs. $V_{ov}$/EOT. Horizontal dashed lines mark the true $\mu$. The $x$-axis is the same as in (i-l). **(i)** The threshold voltage $V_T$ extracted with each method for device families with $\phi_B$ = 0.15 eV, **(j)** 0.3 eV, **(k)** 0.45 eV, and **(l)** 0.6 eV plotted vs. $V_{ov}$/EOT. The horizontal dashed lines mark the true channel $V_T$.

Next, in **Figures 3i-l**, we plot the $V_T$ extracted from our proposed method, the linear extrapolation method, and the Y-function method vs. $V_{ov}$/EOT using the same horizontal $x$-axis limits as in Figures 3e-h. Importantly, in these transistors, contact gating obscures the channel turn-on, causing the linear extrapolation and Y-function methods to significantly overestimate $V_T$. In comparison, we find that our proposed extraction tends to yield much more accurate $V_T$ estimates, with a worst-case $V_T$ error of 0.2 V in the range of $V_{ov}$/EOT plotted in Figures 3i-l. We note that our method and the conventional methods do not always converge to the true $V_T$ at higher $V_{ov}$, but this is acceptable because the error in estimated $V_T$ is less important (i.e., has smaller impact on the predicted charge carrier density) at large $V_{ov}$.

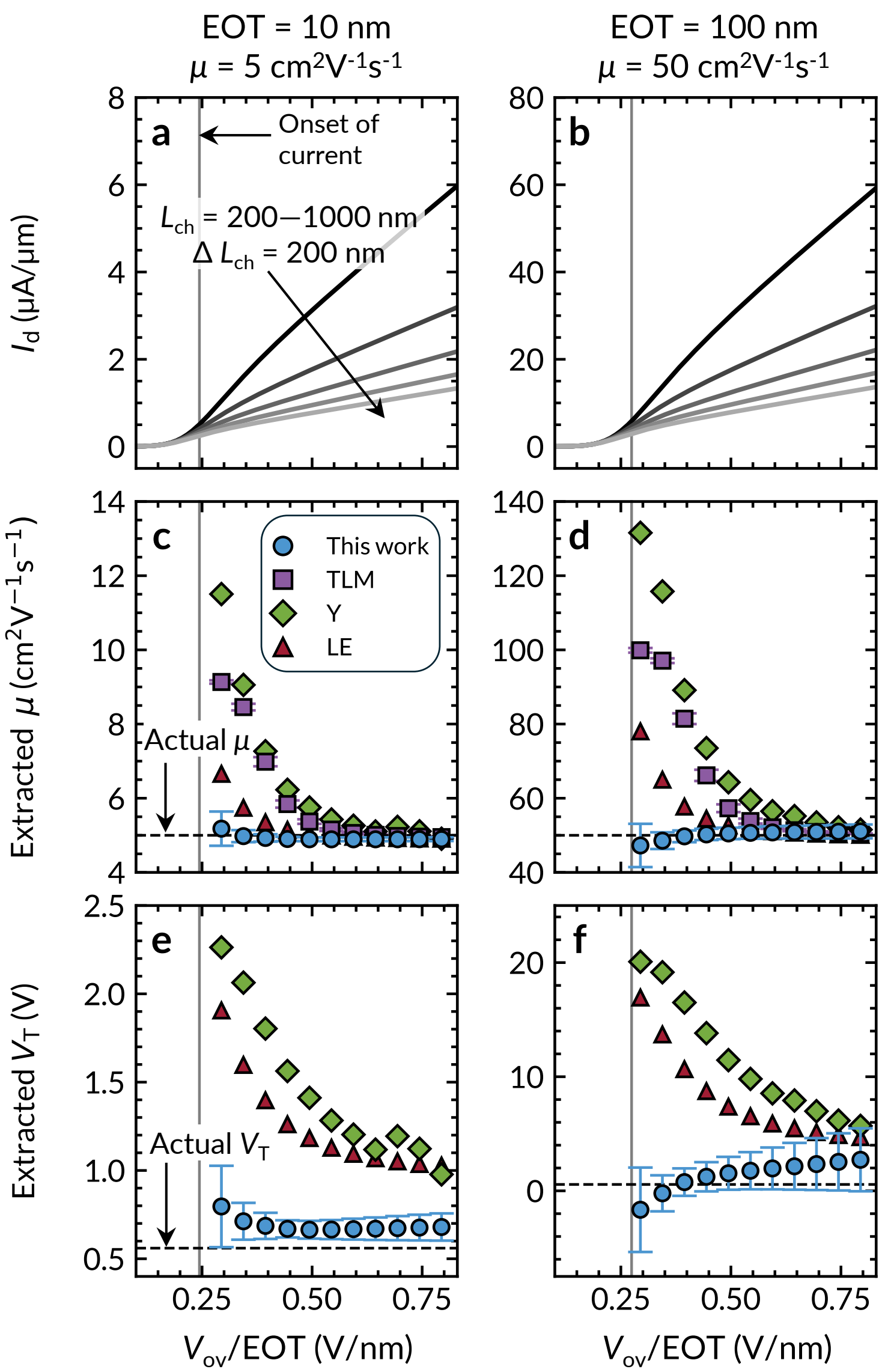

**Figure 4: Validation of our proposed extraction on low-mobility and high-EOT transistors.** $I_d$ vs $V_{ov}$ for families of devices with $L_{ch}$ = 200, 400, ..., 1000 nm and Schottky barrier height $\phi_B$ = 0.45 eV with **(a)** EOT = 10 nm and mobility $\mu$ = 5 cm$^2$V$^{-1}$s$^{-1}$ and **(b)** EOT = 100 nm and $\mu$ = 50 cm$^2$V$^{-1}$s$^{-1}$. The solid gray lines mark the approximate $I_d$ turn-on for each family of devices, and the *x*-axes are the same as in (e,f). **(c)** The extracted mobility $\mu$ with our method and the linear extrapolation (LE), Y-function (Y), and TLM approaches for device families with EOT = 10 nm and mobility $\mu$ = 5 cm$^2$V$^{-1}$s$^{-1}$ and **(d)** EOT = 100 nm and $\mu$ = 50 cm$^2$V$^{-1}$s$^{-1}$ plotted vs. $V_{ov}$/EOT. The horizontal dashed line marks the true $\mu$, and the *x*-axes are the same as in (e,f). **(e)** The threshold voltage $V_T$ extracted with each method for devices with EOT = 10 nm and $\mu$ = 5 cm$^2$V$^{-1}$s$^{-1}$ and **(f)** EOT = 100 nm and $\mu$ = 50 cm$^2$V$^{-1}$s$^{-1}$. Horizontal dashed lines mark the true channel $V_T$.

To ensure that our proposed extraction is applicable to a variety of devices (and not limited to those presented in Figure 3), we repeat similar extractions for back-gated transistors with (i) $\mu$ = 5 $cm^2V^{-1}s^{-1}$ and EOT = 10 nm and (ii) $\mu$ = 50 $cm^2V^{-1}s^{-1}$ and EOT = 100 nm (channel thickness = 0.615 nm and $\phi_B$ = 0.45 eV for all devices). These scenarios are relevant because they correspond to typical devices used to test emerging semiconductor channels. We plot $I_d$ vs. $V_{ov}$ in **Figures 4a,b**, extracted $\mu$ in **Figures 4c,d**, and extracted $V_T$ in **Figures 4e,f**. We find that the trends observed here are similar to those of Figure 3, suggesting that our proposed extraction remains applicable at higher EOTs or lower $\mu$. Thus, the method we propose in this work appears to facilitate accurate extractions from a variety of contact-gated transistors with high $R_C$ and/or early dielectric breakdown that cannot access the higher $V_{ov}$ range necessitated by conventional methods.

**Effect of device-to-device variation:** To assess the robustness of our extraction, we apply it to devices whose $\mu$ and $V_T$ have a certain amount of variation, as would be seen experimentally. For each device, we randomly select $\mu$ and $V_T$ according to Gaussian distributions with means (standard deviations) of 50 $cm^2V^{-1}s^{-1}$ (10%) and 0.56 V (0.1 V), respectively. As before, we use Sentaurus TCAD[18] to generate current-voltage characteristics that we analyze with our proposed method and the TLM approach. Because the Y-function and linear extrapolation methods are applied to one device at a time, they are not affected by variations between devices; thus, we do not re-analyze them here. We quantify each method's accuracy in terms of its mean absolute error (MAE) and its confidence interval coverage probability (CICP; the probability that the true $\mu$ lies within the range of estimated value ± the error). In other words, an MAE near 0% or as small as possible and a CICP close to 100% are desirable. All devices are identical to those used in Figure 3c, i.e., back-gated with EOT = 10 nm, channel thickness = 0.615 nm, and $\phi_B$ = 0.45 eV.

We perform 100 extractions on families of devices with 5 channel lengths ($L_{ch}$ = 200, 400, ..., 1000 nm), starting at high $V_{ov}$/EOT = 0.64 V/nm. We find that our proposed approach and the TLM approach offer reasonably small MAE = 14.3% and 10.9% (on the same order as the $\mu$ standard deviation), respectively, and CICPs of 99% and 65%, respectively (**Figures 5a,b**), indicating that random variation does not significantly affect the accuracy or reliability of these methods at high $V_{ov}$.

Next, we repeat this procedure at smaller $V_{ov}$/EOT = 0.3 V/nm, which lies within the contact-influenced region of the $I_d$ vs. $V_{ov}$ curves in Figure 3c. Here, the MAE of our proposed method increases to 20.2%, and its CICP remains high at 94% (**Figure 5c**). However, the MAE of the TLM approach increases greatly to 116.0% (>2× mobility overestimation), whereas its CICP falls to 0%, i.e., the TLM approach cannot estimate $\mu$ to within error bars across any of the 100 trials (**Figure 5d**). The MAE of our approach can be improved by adding more devices; repeating the extraction at $V_{ov}$/EOT = 0.3 V/nm using three of each $L_{ch}$ (**Figure 5e**;

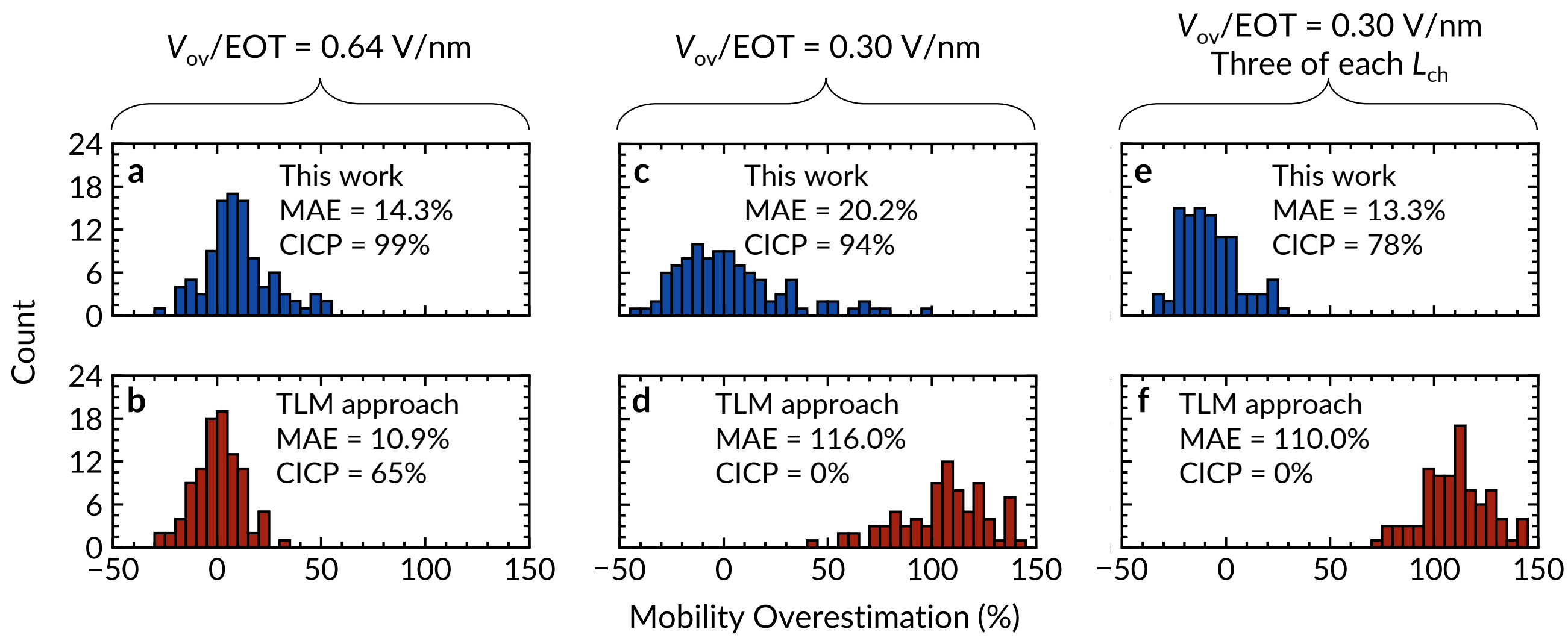

**Figure 5: Robustness to variation of our method and of the TLM approach.** Histograms of mobility overestimations for 100 families of devices with $L_{ch}$ = 200, 400, ..., 1000 nm. The devices are selected at random from a Gaussian distribution with mean (standard deviation) $\mu$ = 50 cm$^2$V$^{-1}$s$^{-1}$ (10%) and mean $V_T$ = 0.56 V (0.1 V). **(a)** Our method and **(b)** the TLM approach at high $V_{ov}$/EOT = 0.64 V/nm. **(c)** Our method and **(d)** the TLM approach at lower $V_{ov}$/EOT = 0.3 V/nm. **(e)** Our method and **(f)** the TLM approach at $V_{ov}$/EOT = 0.3 V/nm with three of each $L_{ch}$ used in the extraction. Each figure lists the mean absolute error (MAE, ideally should be close to 0%) and confidence interval coverage probability (CICP, ideally should be close to 100%).

with devices subject to the same random variations as before) decreases the MAE of our approach to 13.3%, though the CICP also worsens slightly to 78% (which occurs because the estimated standard error shrinks). However, the MAE of the TLM approach only decreases slightly to 110.0% (~2× mobility overestimation) and the CICP remains at 0% (**Figure 5f**), indicating that adding more devices to the TLM analysis is ineffective for improving both accuracy and reliability at $V_{ov}$/EOT = 0.3 V/nm.

We note that although Figure 5c shows our method yields a reasonably low MAE = 20.2% on the entire set of 100 device families, the 10 worst extractions still overestimate $\mu$ by 48% to 102%. However, the CICP considering only these 10 extractions is 100%, i.e., each of these 10 worst extractions also yielded large estimated errors that encompassed the true $\mu$ of 50 cm$^2$V$^{-1}$s$^{-1}$. Thus, although our method can overestimate $\mu$ of contact-gated devices (with $\phi_B$ = 0.45 eV), these overestimations are accompanied by large error bars that clearly indicate when such overestimation occurs.

**Discussion:** We have developed a simple method for extracting the mobility and channel threshold voltage from transistors with gate-voltage-dependent contact resistance. We tested this method by analyzing TCAD-generated current-voltage characteristics and showed it can accurately extract the mobility and threshold voltage when devices are heavily influenced by contact gating, even when conventional methods overestimate the mobility by 2–3×. We also find that the standard error associated with the estimated mobility and threshold voltage tends to accurately reflect the actual uncertainty in the extraction, enabling a high confidence extraction of mobility in regimes where the TLM approach fails. Hence, our method expands the range of overdrive voltages that can be used to estimate mobility and threshold voltage, allowing these quantities to be more accurately determined in emerging semiconductor devices with high contact resistance and/or early dielectric breakdown.

**Data availability statement**: The datasets used and/or analysed during the current study available from the corresponding author on reasonable request.

**Code availability statement**: Code that automates the extraction proposed in this work is included in Supplementary Section 4 and in an online GitHub repository.[19]

**Acknowledgements**: R.K.A.B. acknowledges support from the Stanford Graduate Fellowship (SGF) and the NSERC PGS-D programs. The authors also acknowledge partial support from the SRC SUPREME Center.

**Author contributions**: R.K.A.B and E.P. conceived the research idea. R.K.A.B. derived and verified the extraction with assistance from L.H. and C.C.. R.K.A.B. and E.P. wrote the manuscript with input from all authors. A.J.M. and E.P. supervised the research.

**Competing interests**: The authors declare no competing interests.

## References:

1 Cheng, Z. *et al.* How to report and benchmark emerging field-effect transistors. *Nature Electronics* **5**, 416-423, doi:10.1038/s41928-022-00798-8 (2022).

2 Arutchelvan, G. *et al.* From the metal to the channel: a study of carrier injection through the metal/2D $MoS_2$ interface. *Nanoscale* **9**, 10869-10879, doi:10.1039/C7NR02487H (2017).

3 Prakash, A., Ilatikhameneh, H., Wu, P. & Appenzeller, J. Understanding contact gating in Schottky barrier transistors from 2D channels. *Scientific Reports* **7**, 12596, doi:10.1038/s41598-017-12816-3 (2017).

4 Alharbi, A. & Shahrjerdi, D. Analyzing the Effect of High-k Dielectric-Mediated Doping on Contact Resistance in Top-Gated Monolayer $MoS_2$ Transistors. *IEEE Transactions on Electron Devices* **65**, 4084-4092, doi:10.1109/TED.2018.2866772 (2018).

5 Ber, E., Grady, R. W., Pop, E. & Yalon, E. Uncovering the Different Components of Contact Resistance to Atomically Thin Semiconductors. *Advanced Electronic Materials* **9**, 2201342, doi:10.1002/aelm.202201342 (2023).

6 Nasr, J. R., Schulman, D. S., Sebastian, A., Horn, M. W. & Das, S. Mobility Deception in Nanoscale Transistors: An Untold Contact Story. *Advanced Materials* **31**, 1806020, doi:10.1002/adma.201806020 (2019).

7 McCulloch, I., Salleo, A. & Chabinyc, M. Avoid the kinks when measuring mobility. *Science* **352**, 1521-1522, doi:10.1126/science.aaf9062 (2016).

8 Bittle, E. G., Basham, J. I., Jackson, T. N., Jurchescu, O. D. & Gundlach, D. J. Mobility overestimation due to gated contacts in organic field-effect transistors. *Nature Communications* **7**, 10908, doi:10.1038/ncomms10908 (2016).

9 Pang, C.-S. *et al.* Mobility Extraction in 2D Transition Metal Dichalcogenide Devices—Avoiding Contact Resistance Implicated Overestimation. *Small* **17**, 2100940, doi:10.1002/smll.202100940 (2021).

10 Liu, C. *et al.* Device Physics of Contact Issues for the Overestimation and Underestimation of Carrier Mobility in Field-Effect Transistors. *Physical Review Applied* **8**, 034020, doi:10.1103/PhysRevApplied.8.034020 (2017).

11 Mitta, S. B. *et al.* Electrical characterization of 2D materials-based field-effect transistors. *2D Materials* **8**, 012002, doi:10.1088/2053-1583/abc187 (2021).

12 English, C. D., Shine, G., Dorgan, V. E., Saraswat, K. C. & Pop, E. Improved Contacts to $MoS_2$ Transistors by Ultra-High Vacuum Metal Deposition. *Nano Letters* **16**, 3824-3830, doi:10.1021/acs.nanolett.6b01309 (2016).

13 Ng, H. K. *et al.* Improving carrier mobility in two-dimensional semiconductors with rippled materials. *Nature Electronics* **5**, 489-496, doi:10.1038/s41928-022-00777-z (2022).

14 Wu, P. Mobility overestimation in molybdenum disulfide transistors due to invasive voltage probes. *Nature Electronics* **6**, 836-838, doi:10.1038/s41928-023-01043-6 (2023).

15 Ng, H. K. *et al.* Reply to: Mobility overestimation in molybdenum disulfide transistors due to invasive voltage probes. *Nature Electronics* **6**, 839-841, doi:10.1038/s41928-023-01044-5 (2023).

16 Jin, H., Xing, Z., Yangyuan, W. & Ru, H. New method for extraction of MOSFET parameters. *IEEE Electron Device Letters* **22**, 597-599, doi:10.1109/55.974590 (2001).

17 Chang, H.-Y., Zhu, W. & Akinwande, D. On the mobility and contact resistance evaluation for transistors based on MoS2 or two-dimensional semiconducting atomic crystals. *Applied Physics Letters* **104**, 113504, doi:10.1063/1.4868536 (2014).

18 Synopsys Inc., Sentaurus Device. Sunnyvale CA, USA. (2017).

19 Bennett, R.K.A. *GitHub Repository*. Available online: github.com/RKABennett/VT_mu_extraction.

20 Chiquito, A. J. *et al.* Back-to-back Schottky diodes: the generalization of the diode theory in analysis and extraction of electrical parameters of nanodevices. *Journal of Physics: Condensed Matter* **24**, 225303, doi:10.1088/0953-8984/24/22/225303 (2012).

21 Lundstrom, M. S. *Fundamentals of Nanotransistors*. Chapter 9.4: The Mobile Charge: Extremely Thin SOI; The Mobile Charge Below Threshold. (World Scientific Publishing, 2015).

22 Nipane, A., Teherani, J. T. & Ueda, A. Demystifying the role of channel region in two-dimensional transistors. *Applied Physics Express* **14**, 044003, doi:10.35848/1882-0786/abf0e1 (2021).

23 Vaknin, Y., Dagan, R. & Rosenwaks, Y. Schottky Barrier Height and Image Force Lowering in Monolayer $MoS_2$ Field Effect Transistors. *Nanomaterials* **10**, 2346, doi:10.3390/nano10122346 (2020).

24 Dickinson, R. G. & Pauling, L. THE CRYSTAL STRUCTURE OF MOLYBDENITE. *Journal of the American Chemical Society* **45**, 1466-1471, doi:10.1021/ja01659a020 (1923).

25 Zhang, L. *et al.* Electrochemical Ammonia Synthesis via Nitrogen Reduction Reaction on a MoS2 Catalyst: Theoretical and Experimental Studies. *Advanced Materials* **30**, 1800191, doi:10.1002/adma.201800191 (2018).

26 Das, S. *et al.* Transistors based on two-dimensional materials for future integrated circuits. *Nature Electronics* **4**, 786-799, doi:10.1038/s41928-021-00670-1 (2021).

# Supplementary Information

# Mobility and Threshold Voltage Extraction in Transistors with Gate-Voltage-Dependent Contact Resistance

Robert K. A. Bennett,[1] Lauren Hoang,[1] Connor Cremers,[1] Andrew J. Mannix,[2] and Eric Pop[1,2,3*]

[1]*Department of Electrical Engineering, Stanford University, Stanford, CA 94305, U.S.A.*

[2]*Department of Materials Science and Engineering, Stanford University, Stanford, CA 94305, U.S.A.*

[3]*Department of Applied Physics, Stanford University, Stanford, CA 94305, U.S.A.*

[*]Contact: epop@stanford.edu

## Supplementary Section 1. Mobility and Threshold Voltage Extraction Tutorial with Accompanying Python Code

We include Python code that automates our extraction procedure, along with $I_d$ vs. $V_{ds}$ data for a sample extraction, in a GitHub Repository.[1] The code is also provided in **Supplementary Section 4**. Here, we provide a tutorial to explain key aspects of this code.

### Supplementary Section 1.1. Preliminary Notes and Setting Up

The accompanying code was developed and tested on Python 3.10.12. Running the code requires the following packages: NumPy, statsmodels, and matplotlib (optional, only needed if plotting).

The structure of the code assumes that all $I_d$ vs. $V_{ds}$ sweeps are saved in one central directory. This directory contains one subdirectory for each channel length $L_{ch}$, where we use the naming convention:

```
Lch=X
```

where `X` is the channel length in µm. Each subdirectory contains comma-separated value (csv) files with $I_d$ vs. $V_{ds}$ sweeps at various fixed $V_{gs}$ values using the following naming convention:

```
IdVd_Vgs=Y.csv
```

where `Y` is the fixed $V_{gs}$ value used. Every file contains two column vectors; the first column contains $V_{ds}$ values in units of volts, and the second column contains the corresponding $I_d$ values in units of A/µm. (Here, the current must be normalized by the channel width. Experimental devices must be properly patterned to prevent errors from current spreading.[2]) Files are comma-delimited, and the first line of each file is ignored (so that it may be used as a header).

The Python file `VT_mu_extraction.py` contains the functions used in this procedure, and `sample_extraction.py` is a short script that calls these functions on $I_d$ vs. $V_{ds}$ data (we provide sample data in `IdVd_data.zip` in the GitHub Repository[1]). Before proceeding, you should download both `.py` files and extract `IdVd_data.zip` (or the appropriate data to be analyzed) to the same directory where the `.py` files are saved. Note that the code assumes that (i) all $I_d$ vs. $V_{ds}$ are forward sweeps, i.e., they are sorted in order of ascending $V_{ds}$, and (ii) each file contains only one $I_d$ vs. $V_{ds}$ sweep.

## Supplementary Section 1.2. Running the Code

The central function that implements the extraction procedure outlined in Figure 2 in the main text, `extraction`, is contained in the module `VT_mu_extraction.py` and called in the script `sample_extraction.py`. To perform the extraction, run the script as provided.

In this script, we set the channel lengths, $V_{gs}$ values, and equivalent oxide thickness (in units of µm, V, and nm, respectively):

```
Lchs = [0.2, 0.4, 0.6, 0.8, 1.0]
vgs_vals = [3.1, 3.2, 3.3, 3.4, 3.5]
EOT = 10
```

We then set the value of $I_d^T$ used in the extraction as 1 µA/µm (entered in units of A/µm):

```
IdT = 1e-6
```

The value of $I_d^T$ should be chosen based on your $I_d$ vs. $V_{ds}$ data. As we discuss in the main text, you should choose $I_d^T$ such that all $V_{ds}^{(i)} \ll V_{gs} - V_T$. Typically, we find this condition can be met by taking $I_d^T$ as the $I_d$ for the longest channel device, at the lowest $V_{gs}$ considered, at $V_{ds} \approx 50$ mV. To determine if $I_d^T$ is chosen properly, change it by ~25%. If the extracted mobility $\mu$ or threshold voltage $V_T$ change noticeably, $I_d^T$ is likely too large. Note that the code will throw an error if any $I_d$ vs. $V_{ds}$ sweeps do not reach $I_d = I_d^T$.

The last variable we assign is `NMC`, which is the number of Monte Carlo steps we perform (see **Supplementary Section 2**). We find that typically ~1000 Monte Carlo steps is sufficient. As a rule of thumb, the number of steps is large enough when (i) it is at least 100 and (ii) doubling it does not noticeably change the result of the extraction.

The script uses the following call to assign the extracted $\mu$, the estimated standard error in $\mu$, the extracted $V_T$, and the estimated standard error in $V_T$ (here, standard error is equivalent to the estimated 68% confidence interval) to the variables `mu_Vds`, `mu_Vds_error`, `VT_Vds`, and `VT_Vds_error`, respectively, and then prints the extracted values to the console:

```
mu_Vds, mu_Vds_error, VT_Vds, VT_Vds_error = extraction(
        foldername_Vds,
        Lchs,
        Vgs_vals,
        EOT,
        IDT,
        NMC,
        plot_Vdsi_extractions = True,
        plot_deltaVC_extractions = True,
        plot_histograms = True)
```

If the last three Boolean variables are `True`, the code will generate and save plots of (i) the $L_{ch}$ vs $V_{ds}^{(i)}$ extraction (similar to Figure 2b in the main text) for every $V_{gs}$, (ii) extracted voltage drops across the contacts for every $V_{gs}$ (similar to Figure 2c in the main text), and (iii) histograms of distributions from the Monte Carlo procedure, respectively. We recommend inspecting each of these plots to ensure that data was processed correctly and to ensure all extracted quantities make physical sense.

## Supplementary Section 2. Monte Carlo Approach for Error Propagation

In our proposed extraction described in the main text, we estimate the intrinsic voltages $V_{ds}^{\prime(i)}$ and $V_{gs}^{\prime}$ based on the extracted voltage drop across the contacts, $\Delta V_C$:

$$V_{ds}^{\prime(i)} = V_{ds}^{\prime(i)} - \Delta V_C \quad (1)$$

$$V_{gs}^{\prime} \approx V_{gs} - \frac{3}{4}\Delta V_C \quad (2)$$

Here, any uncertainty in $\Delta V_\mathrm{C}$ will lead to uncertainty in $V_\mathrm{ds}^{\prime(\mathrm{i})}$ and $V_\mathrm{gs}^{\prime}$ that we must propagate throughout the extraction until we eventually apply linear regression to extract $V_\mathrm{T}$ and $\mu$ in Figure 2e in the main text. Problematically, the errors among data in Figure 2e are not statistically independent from one another, whereas many conventional error estimation methods for linear regression require such independence. As conventional error estimation techniques are therefore inappropriate for estimating uncertainty in our $V_\mathrm{T}$ and $\mu$ extraction, we instead use the Monte Carlo approach described here to improve our estimates of the nominal $V_\mathrm{T}$ and $\mu$ and to estimate their uncertainties.

We denote uncertainty in $\Delta V_\mathrm{C}$ as $\sigma_{\Delta V_\mathrm{C}}$. We assume errors in $V_\mathrm{ds}^{(\mathrm{i})}$ in Figure 2b in the main text are randomly distributed, taking $\sigma_{\Delta V_\mathrm{C}}$ as the standard error associated with the $y$-intercept from linear regression (i.e., the 68% confidence interval). The data points in Figure 2e in the main text (from which $V_\mathrm{T}$ and $\mu$ are eventually extracted) have error in both their $x$ and $y$ values; we denote these errors as $\sigma_x$ and $\sigma_y$, respectively. We calculate $\sigma_x$ and $\sigma_y$ using standard error propagation techniques (implemented in Python code that we have uploaded to a GitHub Repository[1] and included in Supplementary Section 4).

Next, we employ a Monte Carlo approach to simulate how error in $\Delta \mathrm{V}_\mathrm{C}$ propagates to an individual trial when extracting $V_\mathrm{T}$ and $\mu$. To do so, we first compile the usual plot in Figure 2e in the main text, denoting the $j^{\mathrm{th}}$ $x$ and $y$ pair as $x^{(j)}$ and $y^{(j)}$. Then, for each $x^{(j)}$ and $y^{(j)}$ pair, we:

1. Calculate $\sigma_{\Delta V_\mathrm{C}}$ for the $x^{(j)}$ and $y^{(j)}$ pair.
2. Calculate $\sigma_x^{(\mathrm{j})}$ and $\sigma_y^{(\mathrm{j})}$ for the $x^{(j)}$ and $y^{(j)}$ pair.
3. Estimate an adjusted $\tilde{x}^{(\mathrm{j})}$ as a number drawn randomly from a Gaussian distribution with mean $x^{(j)}$ and standard deviation $\sigma_x^{(\mathrm{j})}$.
4. Estimate an adjusted $\tilde{y}^{(\mathrm{j})}$ as a number drawn randomly from a Gaussian distribution with mean $y^{(j)}$ and standard deviation $\sigma_y^{(\mathrm{j})}$.

We then estimate $V_\mathrm{T}$ and $\mu$ for the Monte Carlo trial (denoted $\widetilde{V_\mathrm{T}}$ and $\tilde{\mu}$) as we do in the main text in Figure 2e: we plot all new $\tilde{x}^{(\mathrm{j})}$ and $\tilde{y}^{(\mathrm{j})}$ pairs, perform linear regression to find the slope $\widetilde{m}$ and the intercept $\tilde{b}$, and calculate the $V_\mathrm{T}$ and $\mu$ for the Monte Carlo trial as $\widetilde{V_\mathrm{T}} = \widetilde{m}/2$ and $\tilde{\mu} = 2I_\mathrm{d}^\mathrm{T}/(\tilde{b}C_\mathrm{ox}W)$.

We repeat this procedure to simulate ~1000 distributions of $\widetilde{V_\mathrm{T}}$ and $\tilde{\mu}$. To be consistent with Gaussian statistics (where 68% of data in a normal distribution is within one standard deviation of its median), we take the standard deviations as half of the range that encompasses 68% of each distribution (centered on the median values), and we take the nominal values as the centers of these ranges.

## Supplementary Section 3. Effect of $V_T$ Estimation on TLM Approach for Extracting Mobility

In the transfer length method (TLM) approach[3,4], we first measure the $I_d$ vs. $V_{gs}$ characteristics of a family of field-effect transistors (FETs) with different channel lengths ($L_{ch}$) and estimate the threshold voltage ($V_T$) for each $I_d$ vs. $V_{gs}$ sweep. Here, we assume such measurements do not have substantial hysteresis,[4,5] i.e., the change $\Delta V_T$ between the forward and backward voltage sweep $V_T$ is negligible. This could be < 5% of the total overdrive voltage ($V_{ov} = V_{gs} - V_T$) used, i.e., $\Delta V_T/V_{ov} < 0.05$.

Next, we plot $I_d$ vs. $V_{ov}$ for each $L_{ch}$, find the largest $V_{ov}$ accessible for each device (e.g., if limited by early gate dielectric breakdown), and record the $I_d$ for each device at that common maximum $V_{ov}$. Using these $I_d$, we calculate the total resistance $R_{tot} = V_{ds}/I_d$ for each $L_{ch}$ (where $V_{ds}$ is kept small to ensure that the FET operates in the linear regime). We then plot $R_{tot}$ vs. $L_{ch}$ and perform linear regression to find the slope, which is equal to the sheet resistance $R_{sh}$ at that $V_{ov}$. Finally, we calculate mobility as $\mu = 1/(R_{sh}C_{ox}V_{ov})$.

Because $R_{tot}$ is extracted at a common $V_{ov}$, the method used to extract $V_T$ influences the estimated $\mu$. In the main text, we use $V_T$ from linear extrapolation for this purpose (at the highest $V_{gs}$ available, i.e., where the channel is most dominant). Here, we repeat extractions using constant-current threshold voltages[6] $V_{T,CC}$, i.e., the $V_{gs}$ that must be applied to achieve a specified drain current.

We begin by extracting $V_{T,CC}$ for the $I_d$ vs. $V_{gs}$ sweeps in Figures 3a-d in the main text [Schottky barrier height $\phi_B$ = 0.15 to 0.6 eV, $L_{ch}$ = 200 to 1000 nm, equivalent oxide thickness (EOT) = 10 nm, channel thickness = 0.615 nm, $\mu = 50\ cm^2V^{-1}s^{-1}$]. Here, we consider $V_{T,CC}$ at constant currents between 0.1 nA/µm and 100 nA/µm, encompassing the low-power off-current (0.1 nA/µm) and high-performance off-current (10 nA/µm) specified in the 2022 International Roadmap for Devices and Systems (IRDS).[7] We plot these $V_{T,CC}$ in **Supplementary Figure 1**.

Next, we repeat the TLM $\mu$ extraction presented in the main text, this time extracting $R_{tot}$ at a common $V_{ov} = V_{gs} - V_{T,CC}$ using the $V_{T,CC}$ plotted above. For comparison, we also extract $\mu$ from the TLM method using the true channel inversion $V_T$, which is known *a priori* for our TCAD-generated $I_d$ vs. $V_{gs}$ data (but is difficult to determine for experimental contact-gated FETs, as we discuss in the main text).

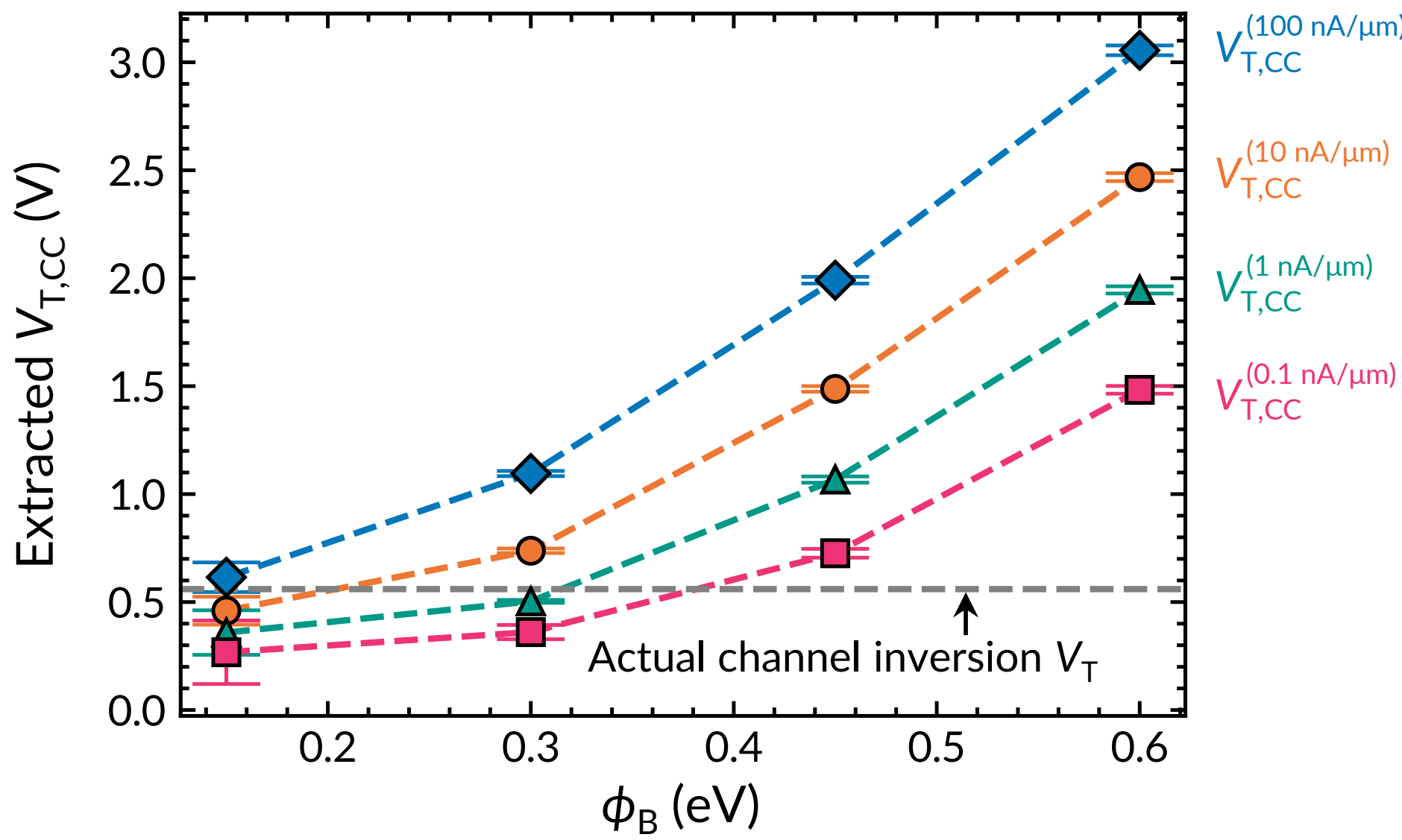

**Supplementary Figure 1: Constant-current threshold voltages $V_{T,CC}$ from $I_d$ vs. $V_{gs}$ curves in Figures 3a-d of the main text.** Error bars show the range of $V_{T,CC}$ extracted from $L_{ch}$ = 200 to 1000 nm. In labels, superscripts denote the current at which $V_{T,CC}$ was extracted. The horizontal dashed line marks the true channel inversion $V_T$.

As shown in **Supplementary Figure 2**, using the true channel inversion $V_T$ allows the TLM approach to accurately extract the $\mu$ with < 10% error even in strongly contact-gated devices. This result suggests that the TLM approach fails in contact-gated devices in the main text due to $V_T$ estimation error propagating to the extracted $\mu$. We also find that when using $V_{T,CC}$, the accuracy of the TLM approach depends on the current at which $V_{T,CC}$ is extracted. When $V_{T,CC}$ happens to be close to the channel inversion $V_T$ (see Supplementary Figure 1), the TLM approach becomes accurate. We also note that the error in the extracted $\mu$ decreases at higher $V_{ov}$, which occurs because errors in $V_T$ have a larger impact on $V_{ov}$ when $V_{ov}$ itself is small. (If the error in $V_T$ is $\Delta V_T$, the relative error in $V_{ov}$ is $\delta V_{ov} = |\Delta V_T / V_{ov}| = |\Delta V_T/(V_{gs} - V_T)|$, which tends to 0 as $V_{gs}$ increases.)

The above analysis suggests that the TLM approach can be suitable when the channel inversion $V_T$ can be accurately determined; however, as we discuss in the main text, extracting this $V_T$ from $I_d$ vs. $V_{gs}$ characteristics of contact-gated devices remains challenging. (Additionally, even with the correct $V_T$, error bars still do not accurately reflect the true uncertainty in the TLM-extracted $\mu$.) Our results indicate that $V_{T,CC}$ taken at 0.1 nA/μm generally allows for the TLM method to accurately extract $\mu$ for the transistors we investigate in this work; however, the current at which $V_{T,CC}$ matches the channel inversion $V_T$ can vary significantly depending on device specifics. Thus, we caution against assigning a universal common $V_{T,CC}$ for $\mu$ extraction using the TLM approach.

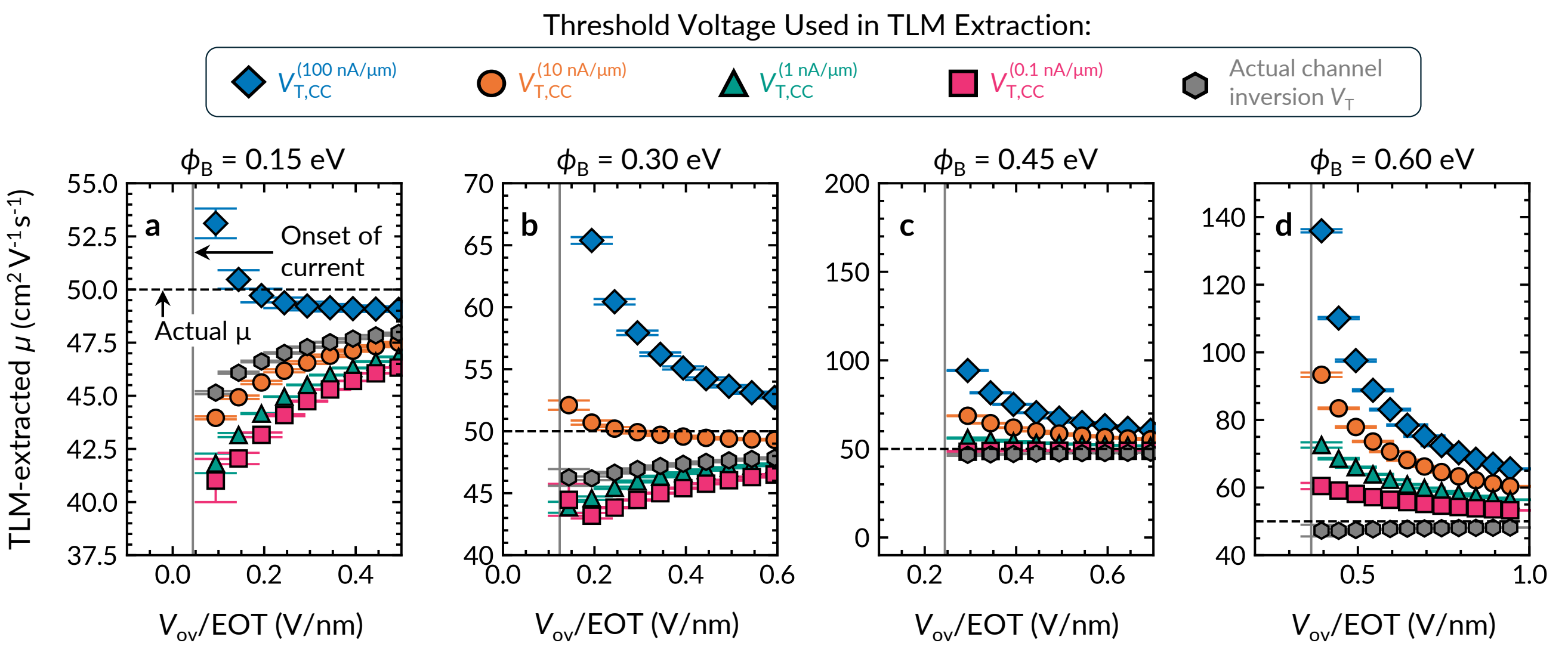

**Supplementary Figure 2: Impact of $V_T$ extraction on TLM approach accuracy.** Mobility $\mu$ extracted using the TLM approach for families of devices with contact Schottky barrier $\phi_B$ = **(a)** 0.15 eV, **(b)** 0.3 eV, **(c)** 0.45 eV, and **(d)** 0.6 eV. We use various extracted constant-current threshold voltages $V_{T,CC}$ and the true channel inversion $V_T$ to benchmark devices at a common overdrive $V_{ov}$ while estimating $R_{tot}$ for the TLM extraction. Different markers indicate which $V_{T,CC}$ was used for this purpose, as denoted in the legend. The horizontal dashed lines mark the actual $\mu$. The gray vertical lines mark the approximate $V_{ov}$ where each family of devices begins to turn on in their $I_d$ vs. $V_{gs}$ characteristics (same as the vertical lines in Figure 3 in the main text).

## Supplementary Section 4. Python Code to Automate the Proposed Extraction

At the end of this Supplementary Information, we provide Python code to automate the extraction proposed in this work. An up-to-date version of this code and a sample dataset for the extraction can also be found in an online GitHub repository.[1] See Supplementary Section 1 for a tutorial on how to run this code and for a description of how the current vs. voltage data should be formatted.

## Supplementary References

1 Bennett, R.K.A. *GitHub Repository*. Available online: github.com/RKABennett/VT_mu_extraction.

2 Köroğlu, Ç. *et al*. Fringe current correction for unpatterned-channel thin-film transistors including contact resistance and velocity saturation effects. In review (2024).

3 Cheng, Z. *et al.* How to report and benchmark emerging field-effect transistors. *Nature Electronics* **5**, 416-423, doi:10.1038/s41928-022-00798-8 (2022).

4 Das, S. *et al.* Transistors based on two-dimensional materials for future integrated circuits. *Nature Electronics* **4**, 786-799, doi:10.1038/s41928-021-00670-1 (2021).

5 Datye, I. *et al*. Reduction of hysteresis in $MoS_2$ transistors using pulsed voltage measurements. *2D Materials* **6**, 011004, doi:10.1088/2053-1583/aae6a1 (2019).

6 Ortiz-Conde, A. *et al*. Revisiting MOSFET threshold voltage extraction methods. *Microelectronics Reliability* **53**, 90-104, doi:10.1016/j.microrel.2012.09.015 (2013).

7 IEEE. International Roadmap for Devices and Systems – More Moore. Accessed: 2024 April 18. Available online: https://irds.ieee.org/editions/2022/more-moore (2022).

```
from math import sqrt
import matplotlib as mpl
import matplotlib.pyplot as plt
import numpy as np
import random
import os
import sys
from VT_mu_extraction import extraction

# get the current working directory
dir_path = os.path.dirname(os.path.abspath(__file__))
sys.path.append(dir_path + '/../')

data = dir_path + '/IdVd_data'            # directory with Id vs. Vds data
Lchs = [0.2, 0.4, 0.6, 0.8, 1.0]          # channel lengths in um
vgs_vals = [3.1, 3.2, 3.3, 3.4, 3.5]      # gate voltages in V
EOT = 10                                  # equivalent oxide thickness in nm

np.random.seed(0) # seed for reproducibility

IDT = 1e-6
NMC = 500

mu, mu_error, VT, VT_error = extraction(
                       data,
                       Lchs,
                       vgs_vals,
                       EOT,
                       IDT,
                       NMC,
                       plot_Vdsi_extractions = True,
                       plot_deltaVC_extractions = True,
                       plot_histograms = True,
                       )

print('Estimated␣mobility:␣{}␣cm^2␣V^s-1␣s^-1'.format(mu))
print('Estimated␣mobility␣error:␣{}␣cm^2␣V^s-1␣s^-1'.format(mu_error))
print('Estimated␣VT:␣{}␣V'.format(VT))
print('Estimated␣VT␣error:␣{}␣V'.format(VT_error))
```

```python
from math import sqrt
import matplotlib as mpl
import matplotlib.pyplot as plt
import numpy as np
import os
import random
from pathlib import Path
import statsmodels.api as sm

# get the current working directory
dir_path = os.path.dirname(os.path.abspath(__file__))

def extraction(
               foldername,
               Lchs,
               Vgs_vals,
               EOT,
               IDT,
               NMC,
               plot_Vdsi_extractions = False,
               plot_deltaVC_extractions = False,
               plot_histograms = False
               ):
    '''
    Applies the method in [CITATION] on Id vs. Vds data to extract the mobility
    and threshold voltage of contact-gated FETs. Mentions of equations,
    figures, variables, etc can be found in the above reference.

    Keyword arguments:
    foldername          -- Directory where Id vs. Vgs sweeps are stored
    Lchs                -- Array-like object listing out channel lengths used
                           (units: um)
    Vgs_vals            -- Array-like object listing out Vgs values used
                            (units: V)
    EOT                 -- Equivalent oxide thickness (units: nm)
    IDT                 -- I_d^T value used for constant-current extraction
                           (units: uA or uA/um (should be the same as the units
                            for Id)
    NMC                 -- Number of Monte Carlo trials used for error prop
                           (Tip: make sure NMC is large enough by
                           increasing it to ensure that doing so does not
                           change the final extracted values)

    Returns:
    mu                  -- Channel mobility (units: cm^2 V^-1 s^-1)
    mu_err              -- Estimated standard error for mu
                           (units: cm^2 V^-1 s^-1)
    VT                  -- Channel inversion threshold voltage (units: V)
    VT_err              -- Estimated standard error for VT (units: V)
    '''

    if plot_Vdsi_extractions or plot_deltaVC_extractions or plot_histograms:
        Path(dir_path + '/plots').mkdir(parents=True, exist_ok=True)
        # plot settings
        mpl.rcParams['lines.linewidth'] = 1.5
        mpl.rcParams['axes.linewidth'] = 0.9
        mpl.rcParams['xtick.major.width'] = 0.6
        mpl.rcParams['ytick.major.width'] = 0.6
        mpl.rcParams['xtick.minor.width'] = 0.3
        mpl.rcParams['ytick.minor.width'] = 0.3
        mpl.rcParams['font.family'] = 'arial'
        plt.rcParams['xtick.minor.visible'] = True
        plt.rcParams['ytick.minor.visible'] = True
        plt.rcParams["xtick.top"] = True
        plt.rcParams["ytick.right"] = True
        mpl.rcParams['xtick.direction'] = 'in'
        mpl.rcParams['ytick.direction'] = 'in'
        plt.rcParams.update({'font.size': 16})

    ##########################################################################
    #
    # Compile a table of Vds' and Vgs' values similar to that of Fig. 2e in the
    # main text
    #
    ##########################################################################

    data_list = []

    # Extract Vds' and Vgs' values for every Vgs
    for Vgs in Vgs_vals:
        Vdsi_list,\
```

```
    Vch_list,\
    Vgsp,\
    error_VC = extract_at_fixed_Vgs(
                                foldername,
                                Lchs,
                                Vgs,
                                IDT,
                                plot_deltaVC_extractions,
                                plot_Vdsi_extractions
                                )

    data_list.append([Vdsi_list, Vch_list, Vgsp, error_VC])

# Now, we iterate through each channel length. We use our extracted Vgsp
# values in conjunction with the appropriate Vdsp values to come up with
# our x and y values, and their accompanying errors, for the final fit.
# Here, x and y values are those plotted on the x- and y-axes in Fig. 2e.

xy_mat = build_data_matrix(Lchs, Vgs_vals, data_list)

###########################################################################
#
# Perform multiple Monte Carlo iterations to simulate distributions of mu
# and VT based on their estimated errors
#
###########################################################################

mu_list, VT_list, x, y, x_error, y_error = [], [], [], [], [], []

for i in range(NMC):
    mu_i, VT_i = MC_step(xy_mat, EOT, IDT)

    mu_list.append(mu_i)
    VT_list.append(VT_i)

###########################################################################
#
# Extract the nominal mobility/VT and their accompanying standard errors
#
###########################################################################

mu_list = np.sort(mu_list)
VT_list = np.sort(VT_list)
x = np.array(x)
y = np.array(y)

# Here, we filter the histograms by taking the median value +- 34% (i.e.,
# the central 68% of the histograms) to stay roughly in line with Gaussian
# statistics, as we discuss in the Supporting Information.

p1 = 16     # exclude the bottom 16% of the histogram
p2 = 100-p1 # exclude the top 16% of the histrogram

mu_filtered = mu_list[
                      int(np.size(mu_list)* p1/100)
                      :
                      int(np.size(mu_list)* p2/100)
                     ]

VT_filtered = VT_list[
                      int(np.size(VT_list)* p1/100)
                      :
                      int(np.size(VT_list)* p2/100)
                     ]

# Uncomment the following lines to plot histograms of mu and VT before
# and after filtering. (It's normal for unfiltered distributions to contain
# extreme outliers. This is one of the reasons why we filter.)

# Calculate and return the final mu and VT values
mu = (mu_filtered[0] + mu_filtered[-1])/2
mu_error = (mu_filtered[-1] - mu_filtered[0])/2
VT = (VT_filtered[0] + VT_filtered[-1])/2
VT_error = (VT_filtered[-1] - VT_filtered[0])/2

if plot_histograms:
    histogram_fit(mu_list, 2, foldername, 'mobility')
    histogram_fit(mu_filtered,2, foldername, 'mobility_filtered')
    histogram_fit(VT_list, 0.2, foldername, 'VT')
    histogram_fit(VT_filtered, 0.2, foldername, 'VT_filtered')
```

```
    return mu, mu_error, VT, VT_error

def extract_at_fixed_Vgs(foldername, Lchs,
                         Vgs, IDT, plot_deltaVC_extractions,
                         plot_Vdsi_extractions):
    '''
    Extracts the voltage drop across the channel, Vch, for an Id vs. Vds sweep
    at a fixed target current = I_d^T.

    Keyword arguments:
    foldername          -- Directory where Id vs. Vgs sweeps are stored
    Lchs                -- Array-like object listing out channel lengths used
                           (units: um)
    Vgs                 -- Vgs value at which Id vs. Vds sweep is performed
    IDT                 -- I_d^T value used for constant-current extraction
                           (units: uA or uA/um (should be the same as the units
                            for Id))


    Returns:
    Vdsi_list           -- List containing the V_ds^(i) value for each Lch
    Vdsip_list          -- List containing V_ds^(i)' values for each Lch
    Vgsp                -- Vgs'
    error_delta_VC      -- Estimated standard error in delta V_C
    '''

    Vdsi_list = find_Vds_list(foldername, Lchs, Vgs, IDT,plot_Vdsi_extractions)
    Lchs = np.array(Lchs, dtype = 'float64')

    ##########################################################################
    #
    # Statistics to find b, m, and the error for b
    #
    ##########################################################################

    x = Lchs
    x = sm.add_constant(x)
    y = Vdsi_list

    delta_VC, m, error_delta_VC, error_m = linear_regression(x,y)
    Vdsip_list = Vdsi_list - delta_VC    # List of V_ds^(i)'
    Vgsp = Vgs - 0.75*delta_VC           # Vgs'

    ##########################################################################
    #
    # Optional: uncomment this section to generate a TLM-like plot (same as
    # in Fig. 2b) for the delta V_C extraction.
    #
    ##########################################################################

    if plot_deltaVC_extractions:
        Path(dir_path + '/plots/deltaVC_extractions').mkdir(
                                                    parents=True,
                                                    exist_ok=True)
        plotname = '/deltaVC_extraction_Vgs={}.png'.format(Vgs)
        x = np.array([0, np.max(Lchs)])
        y = m*x + delta_VC

        fig, ax = plt.subplots(1,1)

        ax.plot(
                  Lchs,
                  Vdsi_list*1000,
                  marker = 'o',
                  color = 'k',
                  ls = 'None'
                  )
        ax.plot(
                  x,
                  y*1000,
                  color = 'r',
                  ls = '--',
                  marker = 'None'
                  )

        ax.set_xlim(0,)
        ax.set_xlabel('$L_\\mathrm{ch} (um)$')
        ax.set_ylabel('$V_{DS}$ (mV) at $I_d = I_\\mathrm{D}^\\mathrm{T}$')
        plt.tight_layout()
        plt.savefig(dir_path + '/{}/{}'.format(
                                            'plots/deltaVC_extractions',
```

```
                                              plotname)
                                              )
        plt.close()

    return(Vdsi_list, Vdsip_list, Vgsp, error_delta_VC)

def find_Vds_list(foldername, Lchs,
                  Vgs, IDT, plot_Vdsi_extraction):
    '''
    Finds the Vds values where Id = some target current for a family of IdVd
    curves at different Lchs.

    Keyword arguments:
    foldername          -- TBA
    Lchs                -- Array-like object listing out channel lengths used
                           (units: um)
    Vgs                 -- Vgs value at which Id vs. Vds sweep is performed
    EOT                 -- Equivalent oxide thickness (units: nm)
    IDT                 -- I_d^T value used for constant-current extraction
                           (units: uA or uA/um (should be the same as the units
                           for Id))

    Returns:
    Vdsi_list           -- List of V_ds^(i) values for each Lch values
    '''

    basefilename = foldername + '/Lch={}/IdVd_Vgs={}.csv'
    if plot_Vdsi_extraction:
        Path(dir_path + '/plots/Vdsi_extractions').mkdir(parents=True,
                                                         exist_ok=True)
        fig, ax = plt.subplots(1,1)

    for Lch in Lchs:
        data = np.loadtxt(
                          basefilename.format(Lch, Vgs),
                          skiprows = 1,
                          delimiter = ','
                          ).T
        Vd = data[0]
        Id = data[1]

    Vdsi_list = []

    for Lch in Lchs:
        data = np.loadtxt(
                          basefilename.format(Lch, Vgs),
                          skiprows = 1,
                          delimiter = ','
                          ).T
        Vd = data[0]
        Id = data[1]
        Vds_target = find_Vds_targ(Vd, Id, IDT)
        Vdsi_list.append(Vds_target)

        if plot_Vdsi_extraction:
            ax.plot(
                    Vd*10**3,
                    Id*10**6,
                    label = Lch
                    )
            ax.plot(
                    Vds_target*10**3,
                    IDT*10**6,
                    marker = 'o',
                    color = 'k',
                    )
            ax.axhline(IDT*10**6, ls = '--', color = 'k')

    if plot_Vdsi_extraction:
        ax.set_ylim(0,2*IDT*10**6)
        ax.set_xlim(np.min(Vd)*10**3, 1.1*np.max(Vdsi_list)*10**3)
        ax.set_xlabel('$V_\\mathrm{ds}$ (mV)')
        ax.set_ylabel('$I_\\mathrm{d}$ (uA/um)')
        ax.legend(loc = 'best', title = '$L_\\mathrm{ch}$')
        plt.title('$V_\\mathrm{gs}$ = ' + str(Vgs) + ' V')
        plt.tight_layout()
        plt.savefig(dir_path + '/{}/Vdsi_extraction_Vgs={}.png'.format(
                                                        'plots/Vdsi_extractions',
                                                         Vgs))
        plt.close()
    return np.array(Vdsi_list)
```

```
def find_Vds_targ(Vds, Id, IDT):
    '''
    Finds the Vds value of a single IdVd where Id = some target current.


    Keyword arguments:
    Vds and Id      -- Array-like objects of V_ds and I_d values used in
                       Id vs. Ids sweep (units: V_ds in V; Id in uA or uA/um)
    EOT             -- Equivalent oxide thickness (units: nm)
    IDT             -- I_d^T value used for constant-current extraction
                       (units: uA or uA/um (should be the same as the units
                       for Id))

    Returns:
    Vds_target      -- Vds value at which Id = target current (same as V_ds^(i)
                       in the main text)

    Note that if Vds_target is between two values (which is usually will be),
    we perform linear extrapolation using the two neighboring datapoints. If
    the Id vs. Vds sweep contains multiple Vds values corresponding to the
    chosen IDT (which can happen if you have multiple sweeps in
    your curve or if you have noise), then we return only the first.
    '''

    for i in range(np.size(Vds) - 1):
        if (
            (Id[i] < IDT and Id[i + 1] > IDT)
            or Id[i] == IDT
            ):
            m = (Id[i+1] - Id[i]) / (Vds[i+1] - Vds[i])
            b = Id[i] - m*Vds[i]
            Vds_target = (IDT - b)/m
            return Vds_target

    raise Exception("Could not find Vdsi value. Was your IDT chosen properly?")

def MC_step(xy_mat, EOT, IDT):
    '''
    Implements a single step of the Monte Carlo method (see Section S2). We
    call this function many times to build up distributions of mu and VT that
    we then process to find better estimates for their nominal values and
    corresponding errors.


    xy_mat:         -- matrix generated using 'build_data_matrix' function
    EOT             -- Equivalent oxide thickness (units: nm)
    IDT             -- I_d^T value used for constant-current extraction
                       (units: uA or uA/um (should be the same as the units
                        for Id))


    Returns:
    mu              -- mobility from a single Monte Carlo step (equivalent to
                       \tilde(mu) in Section S2 of the Supporting Information)
    VT              -- Threshold voltage from a single Monte Carlo step
                       (equivalent to \tilde(VT) in Section S2 of the
                       Supporting Information)
    '''
    # unpack the xy_mat for convenience
    xval        = xy_mat[:,0]
    error_x     = xy_mat[:,1]
    yval        = xy_mat[:,2]
    error_y     = xy_mat[:,3]
    error_b     = xy_mat[:,4]

    # lists that we will populate with x and y values (as in Fig. 2e) for our
    # eventual linear regression.
    x = []
    y = []

    # apply the method described in Section S2 of the supporting information to
    # generate a series of perturbed x and y data points
    for j in range(np.size(xval)):
        error_in_b1 =  np.random.normal(0,error_b[j],1)[0]
        error_in_x = error_in_b1 * abs(error_x[j])
        error_in_b2 =  np.random.normal(0,error_b[j],1)[0]
        error_in_y = error_in_b2 * abs(error_y[j])
        new_xval = xval[j] + error_in_x
        new_yval = yval[j] + error_in_y
        x.append(new_xval)
```

```
        y.append(new_yval)

    b, m, b_error, m_error = linear_regression(x,y)

    Cox = 8.85e-12*3.9/(EOT*10**-9)
    A = (10**4/Cox*2*IDT)

    # extract the mobility and VT for the single Monte Carlo trial
    mu = A/b
    Vth = m/2
    return mu, Vth

def build_data_matrix(Lchs, Vgs_vals, data_list):
    '''
    Builds a matrix containing [xvals, xerrs, yvals, yerrs, error_VCs]
    where:
        xvals           -- nominal x values in Fig. 2b
        xerrs           -- error bars prefactor for x values in Fig. 2b
        yvals           -- nominal y values in Fig. 2b
        yerrs           -- error bars prefactor for y values in Fig. 2b
        error_VC        -- error in delta VC

    Here, xerrs and yerrs are prefactors, i.e., they must be multiplied by
    error_VC to get the true error in the x and y values. Thus, xerr and yerr
    are equivalent to thee quantities described in equations S3 and S4 divided
    by sigma_V_C.

    Keyword arguments:
    Lchs                -- Array-like object listing out channel lengths used
                            (units: um)
    Vgs_vals            -- Array-like object listing out Vgs values used
                             (units: V)
    data_list           -- Matrix-like object built using the
                        'extract_at_fixed_Vgs' function

    Returns:
    xy_mat              -- The matrix described above

    '''
    xy_mat = []
    for j in range(len(Lchs)):
        Lch = float(Lchs[j])
        for i in range(len(Vgs_vals)):
            Vdsi_list, Vch_list, Vgsp, error_b = data_list[i]
            Vdsp = Vch_list[j]

            A = 2*Vgsp*Vdsp
            B = Vdsp**2

            errorA = A*sqrt((3/4 / Vgsp)**2 + (1 / Vdsp)**2)
            errorB = B*sqrt(2*(1 / Vdsp)**2)

            error_x = 1 / Lch
            error_y = np.sqrt(errorA**2 + errorB**2)/Lch

            xval = Vdsp/Lch
            yval = (2*Vgsp*Vdsp - Vdsp**2)/Lch

            xy_mat.append([xval, error_x, yval, error_y, error_b])

    return np.array(xy_mat)

def histogram_fit(vals, step,
                  savefoldername, hist_name):
    '''
    Generates and saves a histogram to visualize mu or VT distributions
    obtained from the Monte Carlo procedure.

    Keyword arguments:
    vals        -- quantities of interest (mu or VT values) whose distribution
                    you want to see
    step        -- step size for bins
    foldername  -- directory to which you wish to save the histogram
    hist_name   -- file name for saving

    Returns:
    N/A
    '''
    Path(dir_path + '/plots/histograms').mkdir(parents=True, exist_ok=True)
    fig, ax = plt.subplots(1,1)
    bins = np.arange(np.min(vals), np.max(vals) + 2*step, step)
```

```
    ax.hist(vals, edgecolor = 'k', color = 'gray', bins = bins)
    ax.set_xlim(np.min(vals), np.max(vals))
    ax.set_title('Min␣=␣{},␣max␣=␣{},␣range␣=␣{}'.format(
                    round(vals[0], 2),
                    round(vals[-1],2),
                    round(vals[-1] - vals[0], 2)
                    )
                    )
    ax.set_xlabel('Value')
    ax.set_ylabel('Count')
    plt.tight_layout()
    plt.savefig(dir_path\
            + '/plots/histograms/histogram_{}.png'.format(hist_name))
    plt.close()

def linear_regression(x, y):
    '''
    Implements linear regression for a collection of x,y data.

    Keyword arguments:
    x           -- Array-like object of independent variable
    y           -- Array-like object of dependent variable

    Returns:
    b           -- y-intercept for line of best fit
    m           -- Slope of line of best fit
    b_error     -- Standard error for y-intercept
    m_error     -- Standard error for slope
    '''
    X = sm.add_constant(x)
    model = sm.OLS(y, X).fit()
    results_summary = model.summary()
    b = model.params[0]
    m = model.params[1]
    b_error = model.bse[0]
    m_error = model.bse[1]

    return b, m, b_error, m_error
```